\documentstyle[pra,aps]{revtex}

\twocolumn 

\begin{document}
\draft
\title{Phase-dependent spectra in a driven two-level atom}
\author{Peng Zhou and S. Swain}
\address{Department of Applied Mathematics and Theoretical Physics, \\
The Queen's University of Belfast, \\
Belfast BT7 1NN, United Kingdom.}
\date{}
\maketitle

\begin{abstract}
We propose a method to observe  phase-dependent spectra in resonance
fluorescence, employing a two-level atom driven by a strong coherent field
and a weak, amplitude-fluctuating field. The spectra are similar to those
which occur in a squeezed vacuum, but avoid the problem of achieving
squeezing over a $4\pi $ solid angle. The system shows other interesting
features, such as pronounced gain without population inversion.
\end{abstract}


One of the most striking features of the interaction of a monochromatically
driven two-level atom interacting with a squeezed vacuum is the existence of
phase-dependent features in resonance fluorescence. The relative heights and
widths of the spectrum triplet vary greatly with the phase of the squeezed
vacuum \cite{CLW,absp}. In addition, some peaks may show subnatural
linewidths. Although these effects were predicted over a decade ago, their
experimental verification remains a major challenge in quantum optics. The
principal difficulty confronting the experimenter is that the squeezed field
modes must occupy the whole $4\pi $ solid angle of space.

In this paper we propose a method of observing phase-dependent resonance
fluorescence spectra which avoids this obstacle. The experiment should be
feasible with current technology. We employ a two-level atom driven by a
strong, coherent laser, and in addition, by a weak, amplitude-fluctuating
field of wide bandwidth which replaces the squeezed vacuum. We find similar
phase-sensitive spectral profiles in resonance fluorescence and probe
absorption to those which occur in the squeezed vacuum. A distinction is
that no subnatural linewidths arise. Nevertheless, the observation of these
features would be an important demonstration of the modification of basic
radiative properties by a phase dependent reservoir. 

The effect of the relative phase of two coherent driving fields on the
transient dynamics of two-level atoms has recently been investigated
experimentally \cite{phasedy}, the measured fluorescence intensity strongly
demonstrating the phase-dependence. Schemes have been proposed using the
phase difference of coherent driving fields to control the quantum
interference between different transition channels, thereby manipulating
spontaneous emission \cite{phaseSE} and the shape of the Autler-Townes
doublet \cite{phaseAT} in multi-level atomic systems. A novel frequency
metrological technique and measurements of quantum correlations have been
demonstrated by means of the phase sensitivity of the rate of two-photon
absorption \cite{phaseTP}. The relative phase of two strong lasers is also
widely employed to control the line shape and rate of multiphoton ionization %
\onlinecite{phaseMI1,phaseMI2} and the final products of chemical reactions 
\cite{phaseCR}.

Very recently, Camparo and Lambropoulos \cite{phaseMI2} have investigated
the effect of phase-diffusion of the fundamental field on 3-photon and
1-photon photoionization, and shown that significant control can be
achieved, even in the presence of a large laser linewidth. Measurements of
the two-photon absorption spectrum from a variety of randomly
amplitude-modulated laser fields, including the real Gaussian field with a
coherent component, have been performed \cite{GAM1}. Various spectral
profiles are observed when the coherent component is in-phase or 90 degrees
out-of-phase with the stochastic component. For the Mollow spectrum, Vemuri 
{\it et al.} \cite{stoch} have recently shown that the presence of a
stochastic field, in addition to a coherent driving field, may give rise to
a dramatic narrowing of the linewidths of all three peaks, as well as
enhancement of the inversionless gain.

Here we study the modification of the Mollow triplet as controlled by the
phase difference between applied coherent and stochastic fields. A single
two-level atom is driven by a coherent field with a constant amplitude $E_{c}
$, and a stochastic field with a randomly fluctuating amplitude $E_{s}(t)$.
The atom is also damped in the usual way by the electromagnetic vacuum. The
frequencies of the atomic transition, of the coherent laser and of the
stochastic field are assumed to be identical for simplicity. The master
equation for the density operator $\rho $ of the system is

\begin{eqnarray}
\dot{\rho}&=&-i\left[H_{a-c}+H_{a-s},\,\rho \right]  \nonumber \\
&& +\gamma (2\sigma_{-}\rho \sigma_{+}-\sigma_{+}\sigma_{-}\rho -\rho
\sigma_{+}\sigma_{-}),  \label{master1}
\end{eqnarray}
where

\begin{eqnarray}
&&H_{a-c}=\frac{\Omega }{2}(\sigma _{+}+\sigma _{-}), \\
&&H_{a-s}=\frac{x(t)}{2}\left[ e^{i\phi }\sigma _{+}+e^{-i\phi }\sigma _{-}%
\right] .
\end{eqnarray}
$H_{a-c}$ and $H_{a-s}$ describe the interaction of the atom with the
coherent field and the stochastic field, respectively, $\gamma $ is the
atomic decay constant, $\phi $ is the relative phase of the two fields, $%
\Omega =2|{\bf d}\cdot {\bf e}E_{c}|/\hbar $ is the Rabi frequency of the
coherent field, and $x(t)=2|{\bf d}\cdot {\bf e}E_{s}(t)|/\hbar $ represents
the stochastic amplitude of the atom/stochastic-field interaction, which is
assumed to be a real Gaussian-Markovian random process with zero mean value
and correlation function \onlinecite{GAM1,GAM,three},

\begin{equation}
\langle x(t)x(t^{\prime })\rangle =D\kappa e^{-\kappa |t-t^{\prime }|},
\label{cor1}
\end{equation}
where $D$ is the strength of the stochastic process and $\kappa $ can be
associated with the bandwidth of the stochastic field. The correlation
function (\ref{cor1}) describes a field undergoing amplitude fluctuations,
which result in a finite laser bandwidth $\kappa $ %
\onlinecite{GAM1,GAM,three}. The field may be generated by modulating the
output of a stabilized tunable ring dye laser external to the laser cavity,
and employed to probe the two-photon absorption spectrum %
\onlinecite{GAM1,GAM} and resonance fluorescence spectrum \cite{three}.

For simplicity, we assume that the intensity of the coherent part is much
greater than that of the stochastic field, and the bandwidth $\kappa $ of
the stochastic field is much greater than the atomic linewidth (in other
words, the correlation time $\kappa ^{-1}$ of the stochastic field is very
short compared to the radiative lifetime $\gamma ^{-1}$ of the atom), {\it %
i.e.},

\begin{equation}
\Omega \gg \sqrt{D\kappa }\hspace{0.5cm}\mbox{and}\hspace{0.5cm}\kappa \gg
\gamma .  \label{limit}
\end{equation}
One can then invoke standard perturbative techniques to eliminate the
stochastic variable $x(t)$ \cite{cmme}. The resultant master equation for
the reduced density operator $\rho $ takes the form

\begin{eqnarray}
\dot{\rho} &=&-i\left[ H_{a-c},\,\rho \right]   \nonumber \\
&&+\gamma \left( 2\sigma _{-}\rho \sigma _{+}-\sigma _{+}\sigma _{-}\rho
-\rho \sigma _{+}\sigma _{-}\right)   \nonumber \\
&&+\left[ \left( \sigma _{z}\rho \sigma _{z}-\rho \right) ,\,\left( \alpha
_{0}\sigma _{+}-\alpha _{0}^{\ast }\sigma _{-}\right) \right]   \nonumber \\
&&+\left[ \left( e^{-i\phi }\sigma _{-}+e^{i\phi }\sigma _{+}\right) ,[\rho
,\left( \alpha e^{-i\phi }\sigma _{-}+\alpha ^{\ast }e^{i\phi }\sigma
_{+}\right) ]\right]   \label{CMME}
\end{eqnarray}
with 
\begin{eqnarray}
&&\alpha _{0}=\frac{-iD\Omega \kappa }{8\left( \kappa ^{2}+\Omega
^{2}\right) }\left( 1-e^{i2\phi }\right) ,  \nonumber \\
&&\alpha =\frac{D}{8}\left[1+e^{i2\phi} +\frac{\kappa^{2}}
{\kappa ^{2}+\Omega ^{2}} \left(1-e^{i2\phi} \right)\right] .
\end{eqnarray}

The first and second terms in the right-hand side of the reduced master
equation (\ref{CMME}) describe respectively atomic stimulated transitions by
the coherent field and spontaneous emission by the vacuum, whereas the other
terms are associated with the effect of the stochastic amplitude-fluctuating
field, and are phase dependent. It may be shown that the third term acts
like a driving field contribution, and the last a reservoir-related one. In
other words, the weak, wide bandwidth and amplitude-fluctuating field
effectively forms a phase-dependent reservoir.

This becomes clearer if we rewrite the equation (\ref{CMME}) for $\phi =0$
as,

\begin{eqnarray}
\dot{\rho} &=&-i\left[ H_{a-c},\,\rho \right]   \nonumber \\
&&+\gamma (N+1)\left( 2\sigma _{-}\rho \sigma _{+}-\sigma _{+}\sigma
_{-}\rho -\rho \sigma _{+}\sigma _{-}\right)   \nonumber \\
&&+\gamma N\left( 2\sigma _{+}\rho \sigma _{-}-\sigma _{-}\sigma _{+}\rho
-\rho \sigma _{-}\sigma _{+}\right)   \nonumber \\
&&+2\gamma M\sigma _{+}\rho \sigma _{+}+2\gamma M\sigma _{-}\rho \sigma _{-},
\label{CMME1}
\end{eqnarray}
where $M=N=D/4\gamma $. The master equation (\ref{CMME1}) is the same as
that of a coherently driven two-level atom damped by a reservoir in which
there is the maximal {\it classical} correlation between pairs of photons, a
nonideal squeezed vacuum \cite{CLW} for example. Such a reservoir is
sometimes called a ``classically squeezed field'' (CSF). 

We wish to compare this response to that when the stochastic field is
replaced by an ideal squeezed vacuum (ISV), where $\left| M\right| =\sqrt{%
N(N+1)}$. The interaction of a two-level atom with the ISV is characterized
by the phase $\Phi ,$ the difference between the phases of the driving laser
and the squeezed vacuum \cite{CLW}. The case $\phi =0$ corresponds to $\Phi
=\pi .$

For $\phi =\pi /2,$ we again obtain an equation of the form (\ref{CMME1}),
but this is now identical to the equation for a two-level atom interacting
with a CSF with $\Phi =0.$ In this case $N=-M=D/4\gamma \times \kappa
^{2}/\left( \Omega ^{2}+\kappa ^{2}\right) .$ For $\phi \neq 0,\pi /2$ 
equation (\ref{CMME1}) does not correspond exactly to those for a
classically squeezed field.

In Figure 1 we present the resonance fluorescence spectra for the stochastic
system with $\Omega =200\gamma ,\,\kappa =100\gamma $ and$\,D=10\gamma $ in
frames (a) and (b) for $\phi =0$ and $\pi /2$ respectively, where the strong
phase dependence is evident. In frames (c) and (d) we give the spectra for
the corresponding ideal squeezed vacuum, with $\Omega =20$ and $N=0.25$ in
(c), and $N=0.05$ in (d). (For the squeezed vacuum case we have divided the
parameters by a factor of ten, in order to obtain experimentally reasonable
values for $N.)$ The comparison between the spectra for the stochastic
system and the system with a squeezed vacuum is striking.

It is the modification of the vacuum reservoir by the weak,
amplitude-fluctuating field that strongly affects the physical properties of
the atom. For example, the two quadratures, $\sigma _{x}=\sigma _{-}+\sigma
_{+}$ and $\sigma _{y}=i(\sigma _{-}-\sigma _{+})$, of the atomic
polarization decay at the different rates

\begin{eqnarray}
&&\gamma _{x}=\gamma +\frac{D\kappa ^{2}}{\kappa ^{2}+\Omega ^{2}}\sin
^{2}\phi ,  \nonumber \\
&&\gamma _{y}=\gamma +D\cos ^{2}\phi ,
\end{eqnarray}
whilst the population inversion $\sigma _{z}$ decays at the rate $\gamma
_{z}=\gamma _{x}+\gamma _{y}$. All these decay rates are dependent upon the
relative phase and intensities of the driving fields. Clearly, when the
coherent field is in-phase with the amplitude-diffusing one, {\it i.e.} when 
$\phi =0$, the decay of the dipole quadrature $\sigma _{x}$ is suppressed ($%
\gamma _{x}=\gamma $), while the other decay rate is enhanced ($\gamma
_{y}=\gamma +D$). When both the fields are $\pi /2$ degrees out of phase,
however, the situation is reversed. The suppression or enhancement of the
polarization decays for certain phases may give rise to rich spectral
features.

In Figure 2 we show the global variation of the resonance fluorescence
spectra with the phase $\phi $ for the stochastic system with  $\Omega
=200\gamma ,\,\kappa =100\gamma ,\,D=40\gamma .$ The strong phase dependence
is clear. When  $\phi =0$, the central peak is much narrower and higher than
the sidebands. As the phase increases, the centre peak broadens and
decreases, while the sidebands grow and narrow. For $\phi =\pi /2$, the
central peak has the minimal height and the maximal linewidth, whereas the
sidebands have the opposite characteristics. This behaviour is qualitatively
similar to that of a two-level atom in a squeezed vacuum \cite{CLW}. The
only essential difference is that there are no subnatural linewidths
involved. 

The physics associated with the modification of the Mollow fluorescence
spectrum can be explored by working in the semiclassical dressed states
basis $|\pm \rangle =(|0\rangle \pm |1\rangle )/\sqrt{2}$, which are the
eigenstates of $H_{a-c}$. The condition (\ref{limit}) ensures the secular
approximation to be valid, and consequently, the equations of motion
simplify to

\begin{eqnarray}
&&\dot{\rho}_{z}=-\Gamma _{\parallel }\rho _{z},  \nonumber \\
&&\dot{\rho}_{+-}=-\left( \Gamma _{\perp }+i\Omega ^{\prime }\right) \rho
_{+-},  \label{dbloch}
\end{eqnarray}
where 
\begin{eqnarray}
&&\Gamma _{\parallel }=\gamma _{x},\hspace{0.5cm}\Gamma _{\perp }=\frac{1}{2}%
(\gamma _{y}+\gamma _{z}),  \nonumber \\
&&\Omega ^{\prime }=\Omega \left[ 1+\frac{D\kappa }{2\left( \kappa
^{2}+\Omega ^{2}\right) }\sin ^{2}\phi \right] .
\end{eqnarray}
$\rho _{z}=\left( \rho _{++}-\rho _{--}\right) $ and $\rho _{+-}=\langle
+|\rho |-\rangle $ are the dressed-state inversion and polarization, which
have phase-dependent decay rates $\Gamma _{\parallel }$ and $\Gamma _{\perp }
$, respectively, and $\Omega ^{\prime }$ is the generalized Rabi frequency.
The quantity $\Omega ^{\prime }-\Omega $ represents a dynamical frequency
shift owing to the additional amplitude-fluctuating field. (In fact, it is
consistent with our approximation (\ref{limit}) to set $\Omega ^{\prime
}=\Omega .)$ Obviously, the impact of the weak stochastic field on the
coherently driven atom is merged in the decay rates and level shifts of the
dressed states, which depend on the relative phase of the coherent and
stochastic fields, the correlation strength $D$ and bandwidth $\kappa $ of
the stochastic field.

It follows from (\ref{dbloch}) that the dressed states $|\pm \rangle $ have
the same population. As a result, the resonance fluorescence spectrum is
symmetric:

\begin{equation}
F(\omega )=\frac{\Gamma _{\perp }/8}{\Gamma _{\perp }^{2}+\left( \omega
+\Omega ^{\prime }\right) ^{2}}+\frac{\Gamma _{\parallel }/4}{\Gamma
_{\parallel }^{2}+\omega ^{2}}+\frac{\Gamma _{\perp }/8}{\Gamma _{\perp
}^{2}+\left( \omega -\Omega ^{\prime }\right) ^{2}}.
\end{equation}
The decay of the inversion results in the central peak with linewidth $%
2\Gamma _{\parallel }$ and height $1/4\Gamma _{\parallel }$, whereas the
modulated decays of the coherences give rise to the sidebands with linewidth 
$2\Gamma _{\perp }$ and height $1/8\Gamma _{\perp }$. When $\phi =0$, we
have $\Gamma _{\parallel }=\gamma $ and $\Gamma _{\perp }=3\gamma /2+D$. The
central spectrum is therefore much narrower and higher the sidebands for $%
D\gg \gamma $. When $\phi =\pi /2$, however, $\Gamma _{\parallel }=\gamma
+D\kappa ^{2}/\left( \kappa ^{2}+\Omega ^{2}\right) $ and $\Gamma _{\perp
}=3\gamma /2+\frac{1}{2}D\kappa ^{2}/\left( \kappa ^{2}+\Omega ^{2}\right) $%
. Accordingly, the centre peak widens while the sidebands narrow as the
phase varies from $0$ to $\pi /2$.

The probe absorption spectrum of such a driven atom is also phase dependent. 
We show the spectrum in Fig. 3 with the parameters: $\Omega
=400\gamma ,\,\kappa =100\gamma $ and$\,D=40\gamma $. It is clear that the
absorption spectra are symmetric only when $\phi =0$ or $\,\pi /2$, when the
central component exhibits a Lorentzian lineshape, whilst the sidebands show
the Rayleigh-wing lineshape. Furthermore, there is a very sharp peak at line
centre when $\phi =0$. In these two case, the spectra are qualitatively
similar to those of the corresponding ISV \cite{absp}. 

However, for other values of $\phi $ the spectra are asymmetric, and all
three resonances have Lorentzian-like lineshapes. A striking feature is the
sharp amplification at line centre which occurs in frames (b) and (c). This
amplification takes place without the aid of population inversion. (Note
that the population can never be inverted in either the bare- or
dressed-state basis.) In this respect, the stochastic system differs
markedly from the corresponding CSF or ISV, as neither of these systems show
strong features at line centre. They display only an insignificant `glitch'
at this position. The stochastic system therefore demonstrate some
interesting properties in its own right. A global view of the variation of
the probe absorption with phase is presented in Figure 4.

In conclusion, we have reported a scheme to modify the Mollow fluorescence
and absorption spectra by means of the relative phase of a coherent field
and a stochastically amplitude-diffusing field interacting with a two-level
atom. The phase-sensitive spectral features, which are qualitatively similar
to those of a driven atom in a squeezed vacuum, are revealed. In addition,
we find some new features of this system, for $\phi \neq 0,\pi ,$ including
inversionless gain. Noting that relevant experiments of the phase-control of
the two-photon excitation spectrum of atoms by a field with coherent and
real Gaussian components \cite{GAM1}, and of the transient dynamics of
bichromatically driven two-level atoms \cite{phasedy} have already been
demonstrated, the present model is experimentally accessible. It should be
possible to observe phase dependent spectra without the necessity to squeeze
all $4\pi $ modes of space. Experiments would also demonstrate our ability
to tailor reservoirs so as to modify atomic radiative properties in
fundamental ways. \newline

This work is supported by the United Kingdom EPSRC.

\begin{figure}[tbp]
\caption{Resonance fluorescence spectrum $F(\omega)$ for $\Omega=200\gamma,\,
\kappa=100\gamma,\, D=10\gamma$,  with (a) $\phi=0$  and (b) $\phi=\pi/2$. The
other frames are for the atom in an ideal squeezed vacuum with 
$\Omega=20\gamma$ and $ N=0.25,\, \Phi=\pi$ in (c) and $N=0.05,\,
\Phi=0$ in (d). (All parameters are scaled in units of $\gamma$ throughout 
these figures.)}
\label{fig1}
\end{figure}

\begin{figure}[tbp]
\caption{Three-dimensional fluorescence spectrum $F(\omega)$ against the scaled
frequency $\omega/\gamma$ and the relative phase $\phi/\pi$, for $
\Omega=200\gamma,\, \kappa=100\gamma,\, D=40\gamma$. }
\label{fig2}
\end{figure}

\begin{figure}[tbp]
\caption{Probe absorption spectrum $A(\omega)$, with $\Omega=400\gamma,\,
\kappa=100\gamma,\, D=40\gamma$, and $\phi=0$ in (a), $\phi=\pi/6$ in (b), $%
\phi=\pi/4$ in (c) and $\phi=\pi/2$ in (d). }
\label{fig3}
\end{figure}

\begin{figure}[tbp]
\caption{Same as FIG. 2, but the probe absorption spectrum $A(\omega)$.}
\label{fig4}
\end{figure}

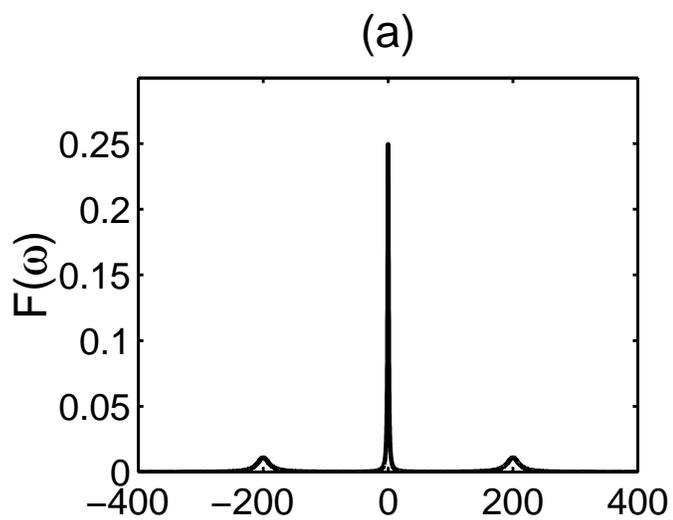
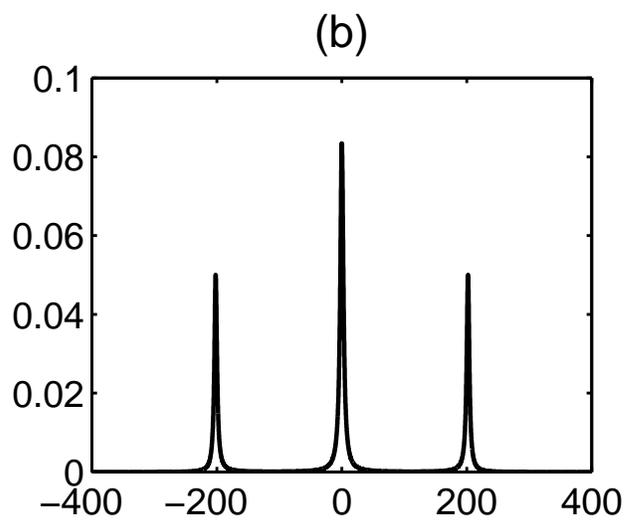
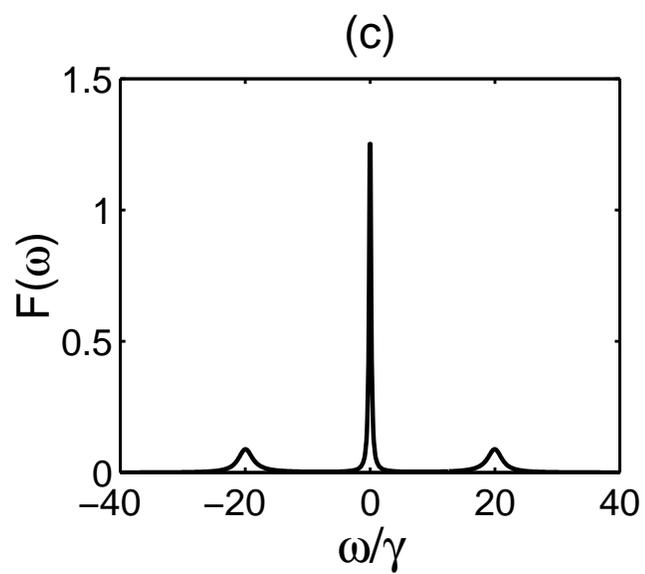
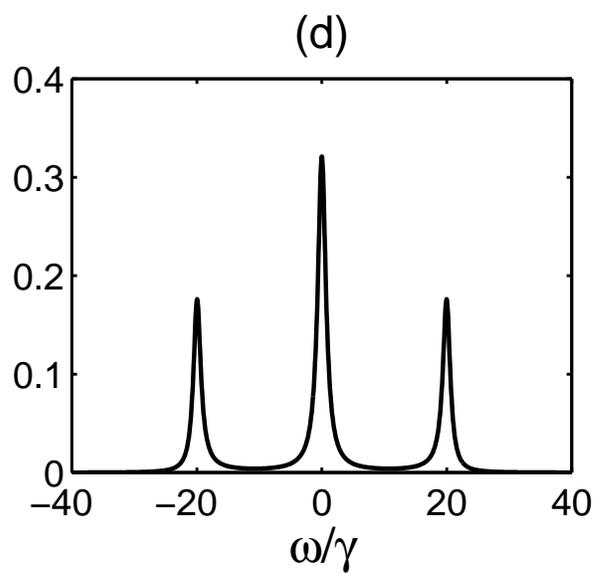
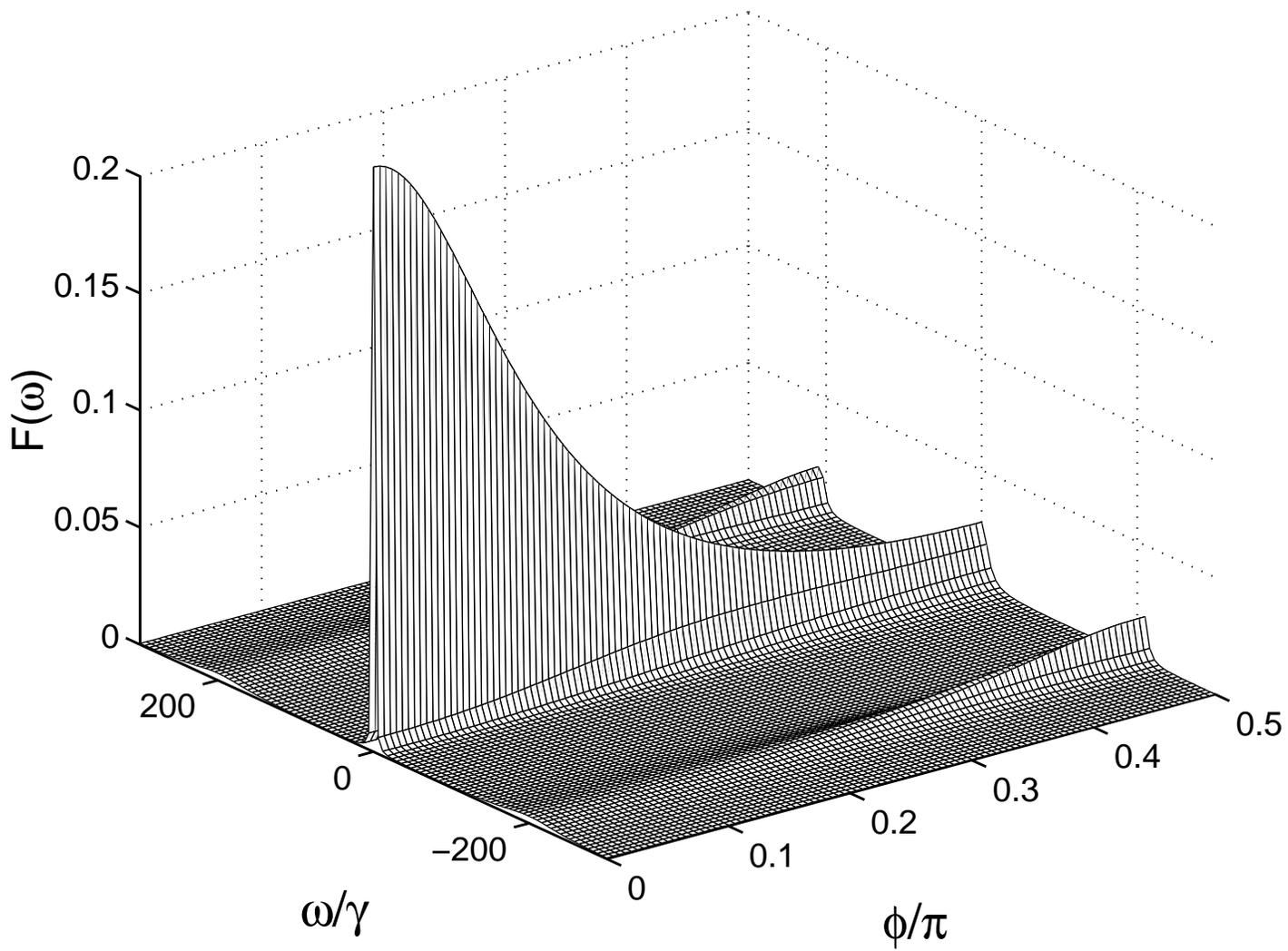
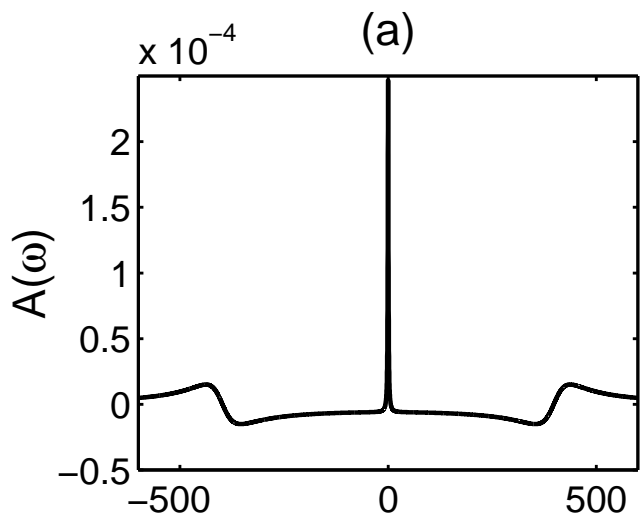
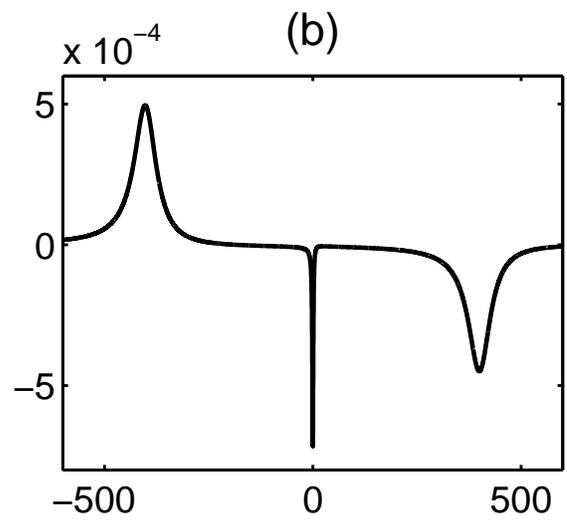
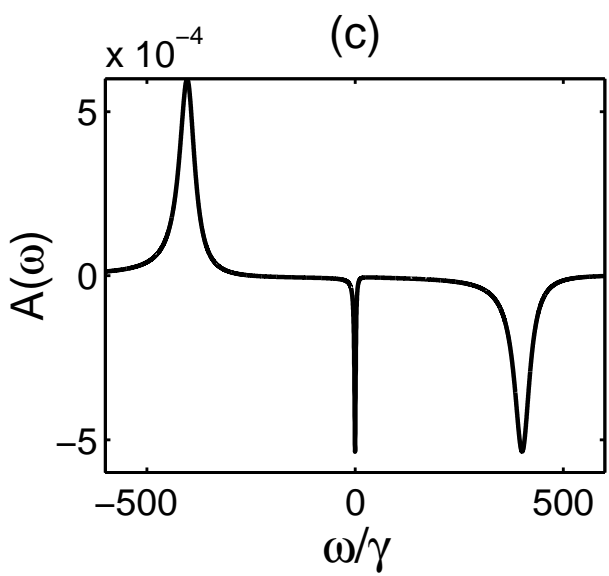
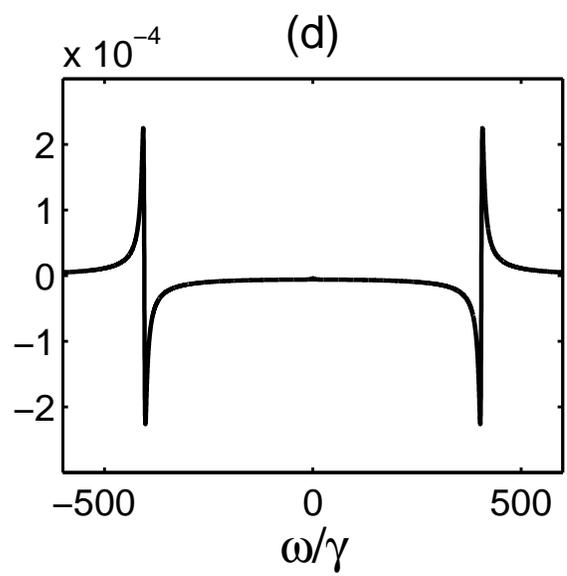
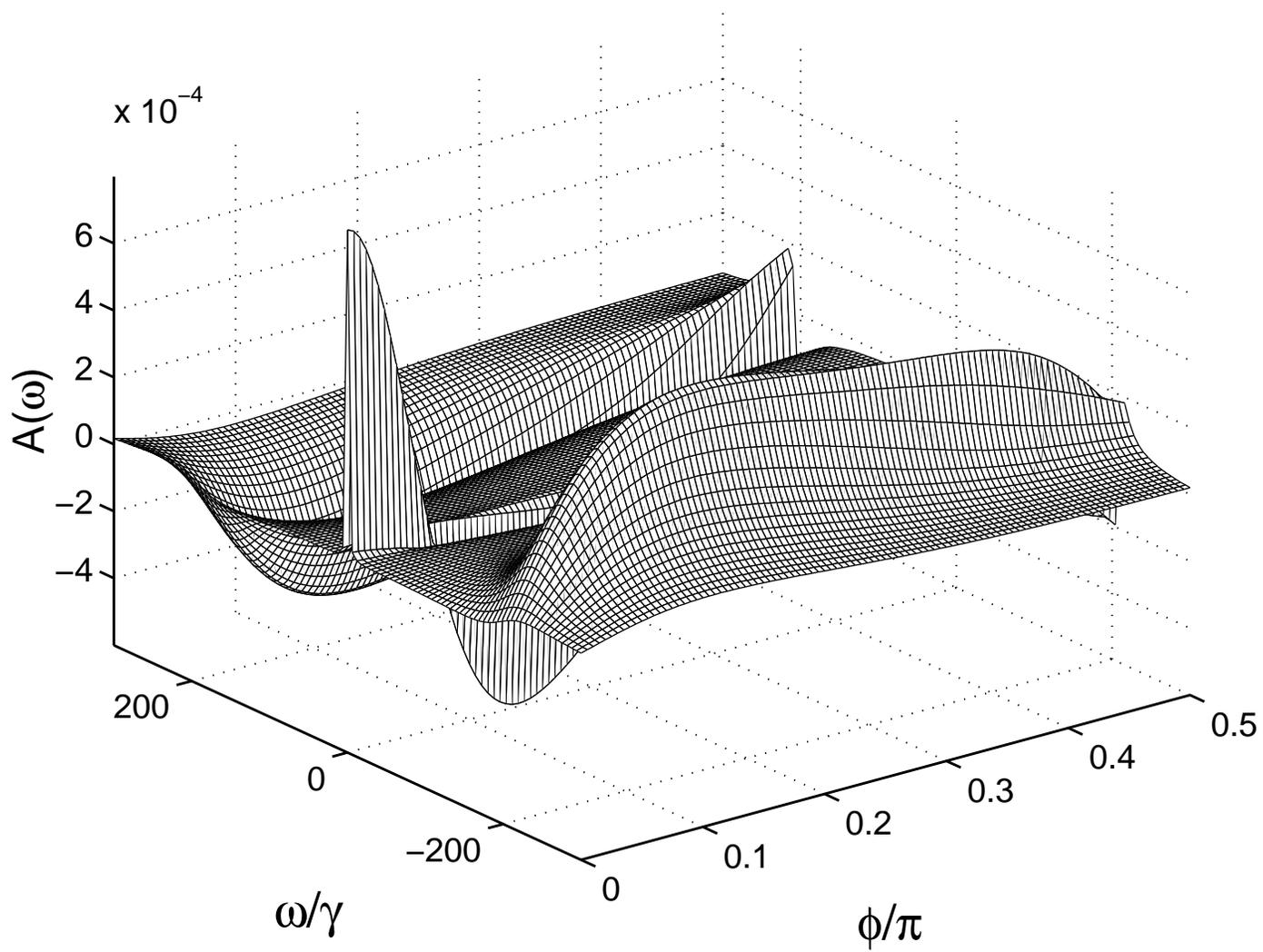
\end{document}